\documentstyle[epsf,aps,preprint]{revtex}

\begin{document}
\title{Rare earth contributions to the X-ray magnetic circular dichroism at
the Co K edge in rare earth-cobalt compounds investigated by multiple-scattering
calculations}
\author{J.P.~Rueff\cite{address}, R.M.~Gal\'era}
\address{Laboratoire de Magn\'etisme Louis N\'eel,\\
associ\'{e} \`{a} l'Universit\'{e} Joseph Fourier de Grenoble,\\
Centre National de la Recherche Scientifique,\\
BP 166X, 38042 Grenoble Cedex, France}
\author{Ch.~Giorgetti, E.~Dartyge}
\address{Laboratoire pour l'Utilisation du Rayonnement Electromagn\'etique, \\
B\^atiment 209D, Centre Universitaire Paris-Sud, 91405 Orsay Cedex, France}
\author{Ch.~Brouder}
\address{Laboratoire de Min\'eralogie-Cristallographie, Universit\'e Paris 6\\
et 7, 4 place Jussieu 75572 Paris Cedex 05, France}
\author{M.~Alouani}
\address{IPCMS-GEMME, Universit\'e Louis Pasteur, 67070 Strasbourg Cedex, France}
\maketitle

\begin{abstract}
The X-ray magnetic circular dichroism (XMCD) has been measured at the Co K
edge in hcp-Co and R-Co compounds (R=La, Tb, Dy). The structure of the
experimental XMCD spectra in the near-edge region has been observed to be
highly sensitive to the magnetic environment of the absorbing site.
Calculations of the XMCD have been carried out at the Co K edge in Co metal,
LaCo$_5$ and TbCo$_5$ within the multiple-scattering framework including the
spin-orbit coupling. In the three systems, the XMCD spectra in the near-edge
region are well reproduced. The possibility to separate the local effects
from those due to the neighboring atoms in the
XMCD cross section makes possible a more physical understanding of the
spectra. The present results emphasize the major role played by the $d$
states of the Tb ions in the XMCD spectrum at the Co K edge in the TbCo$_5$
compound.
\end{abstract}



\section{introduction}

The X-ray magnetic circular dichroism (XMCD) is now well established as a
selective probe of the magnetic properties even in complex systems. XMCD can
yield local quantitative information through a wide range of powerful
calculation techniques. For instance within the multiplet framework, the
XMCD spectra at the L$_{2,3}$ transition metal (TM) and M$_{4,5}$ rare earth
(R) edges can be accurately reproduced \cite{Goedkoop88}. It allows one to
evaluate the local transition metal 3$d$ and rare earth 4$f$ magnetic
moments. Moreover, the sum rules \cite{Thole92,Carra93} can provide, in
favorable cases, a separate determination of both orbital and spin magnetic
moments. However these atomic approaches fail at the TM K edge, where the
excited electron is promoted to the delocalized $p$ states.

For final delocalized band states, the scattering processes encountered by
the photoelectron can be treated in a one-electron picture within the
multiple-scattering framework. This theory was first applied to the
calculation of the x-ray absorption  spectra, and recently it has been
implemented to calculate the XMCD spectra, in particular at the transition
metal K edge. Since the XMCD arises from the interplay between the
spin-polarization and the spin-orbit coupling, these effects have to be
included in the model. Ebert {\it et al.\/} \cite
{Ebert88,Strange90,Stahler93} used a fully relativistic multiple-scattering
formalism  and obtained good results in the near-edge region. This formalism
was recently extended to the EXAFS range \cite{Ebert98}. On
the other hand Ankudinov and Rehr, using a multiple-scattering approach
where the scattering by the atomic sites is fully relativistic and the
propagation between sites is non-relativistic, calculated the
extended structure in XMCD at the L$_{2,3}$ edges of Gd in good agreement
with experiments \cite{Ankudinov95}. However, in this case, the edge region
was not well reproduced. More recently we performed calculations including
both the near-edge and EXAFS regions in the same multiple-scattering
formalism with application to the Fe K edge \cite{Brouder96}. We showed that
the spin-polarization of the $d$-states of the neighboring ions highly
contributes to the signal structure and accounts for the first positive peak
in the Fe XMCD spectrum. This contribution arises from the scattering of the
photoelectron due to the spin-orbit potential of the neighbors and the
absorber. The role of the spin-orbit interaction of the 3$d$ states on the
neighboring sites was also reported by Igarashi and Hirai \cite
{Igarashi96a,Igarashi96b} using tight-binding calculations, and later on by
Guo within an itinerant electronic model \cite{Guo96}. These authors showed
that the XMCD at the TM K edge in Fe and Ni is determined by the orbital
moment on the 3$d$ states through the $p$-$d$ hybridization. By including
the Coulomb interaction between the 3$d$ electrons, the XMCD spectra seem to
be quite well reproduced.

If these techniques now offer a reasonable description of the dichroic
spectrum in pure metals, the interpretation of the XMCD at the TM K edge in
rare earth-transition metal compounds is still open. In particular, the
XMCD signal in the near-edge region may present very different features with
respect to the allied rare earth. Recent experiments performed at the Fe K
edge in R$_2$Fe$_{14}$B compounds were analyzed considering the XMCD
spectra as the addition of the magnetic contributions of both Fe and R
sublattices \cite{Chaboy96}. It was proposed that Fe K edge XMCD is induced
by the splitting of the 5$d$ band, due to the exchange interaction with the 4$f$
states, through the $p$-$d$ hybridization.

In the course of our studies at the L$_{2,3}$ rare earth edges in GdNi$_5$
and TbNi$_5$ \cite{Galera95}, non negligible XMCD signals compared to Ni
metal have been observed at the Ni K edge whereas Ni does not satisfy the
Stoner criterion in the RNi$_5$ compounds \cite{Franse93}. These results
give information on the rare earth influence on the dichroic spectra of the
K edge of Ni. However, the understanding of the XMCD at the K edge of the
transition metal in R-TM intermetallics is far from being achieved and a
better description of the rare earth role has yet to be given. \newline

In this paper we report a comparative study of the Co K edge XMCD spectra
between Co metal and RCo$_x$ metallic alloys with magnetic and non magnetic
rare earth. In most of these compounds, the Co has a strong magnetic moment,
 in the RCo$_5$ compounds for instance, it is comparable to
that of Co metal. The choice of different rare earths allows us to observe
the sensitivity of the XMCD spectrum to the magnetic and chemical
environment of the absorbing site. For Co metal, LaCo$_5$ and TbCo$_5$,
calculations of the XMCD signal have been carried out within the
multiple-scattering model using a semi-relativistic approach
\cite{Brouder96}. These calculations require the knowledge of the atomic
potential which was obtained within the local density approximation using
the linear muffin-tin orbital (LMTO) basis-set within the atomic sphere
approximation and including the so called combined correction term
\cite{Andersen75}. 

The second section of the present paper deals with the XMCD experimental
results observed at the Co K edge in Co metal and various R-Co compounds.
The third section presents the calculations of the crystal  potentials by
means of the linear muffin-tin orbital method within the local-density
approximation, and the multiple-scattering calculations of the x-ray
absorption and the dichroic signal of Co metal, LaCo$_5$ and TbCo$_5$. The
last section summarizes and concludes our paper.

\section{experimental results}

\subsection{samples properties}

Like most of the rare earth-transition metal binary alloys, the R-Co phase
diagrams present a large number of well-defined compounds. The different
crystallographic structures of these compounds are based on the
CaCu$_5$-like hexagonal lattice including simple R or Co atom
substitutions along with layers shifts. The magnetism of the rare
earth-transition metal systems has been extensively studied in the past
two decades (see for instance Refs.\ \cite{Barbara88,Gignoux95}). It
relies on a combination of 4$f$ localized and 3$d$ itinerant magnetism
giving rise to a wide range of magnetic properties, some of them well
suited for technical applications. The band structure of the R-TM metallic
alloys is characterized by the hybridization of the 3$d$ and 5$d$ bands 
\cite{Cyrot79,Shimizu84}. The hybridization leads to a ferromagnetic
coupling between 4$f$ and 3$d$ moments in light rare earth compounds and
ferrimagnetic in heavy rare earth compounds \cite{Campbell72,Brooks89}.
The hybridized states are located at the top of the 3$d$ band and at the
bottom of the 5$d$ band. The Fermi level lies in this region where the
density of states varies strongly. Starting from the pure 3$d$ metal,
the progressive introduction of the rare earth induces a decrease in the
3$d$ magnetism and for given concentrations the Stoner criterion is no
more satisfied. The critical concentration for the onset of the 3$d$
magnetism is RNi$_5$ for the nickel-based alloys whereas it corresponds
to RCo$_2$ for the RCo$_x$ compounds.

The RCo$_2$ compounds crystallize in the MgCu$_2$-type cubic structure
(Laves phases). In YCo$_2$ and LuCo$_2$, Co is non magnetic whereas in
compounds with magnetic rare earths the Co moment of about 1 $\mu_B$, is
induced by exchange interactions between 3$d$ and 4$f$ spins. The Curie
temperature reaches a maximal value of 395 K in GdCo$_2$ and rapidly
decreases with the spin value of the rare earth. On the contrary, in the
RCo$_5$ compounds (CaCu$_5$-like structure) the Co moment, as deduced from
magnetization studies, is approximately 1.64 $\mu_B$ and the strong Co-Co
exchange interaction leads to very high Curie temperatures (T$_C$ $>$ 950
K). The extremely large Co anisotropy favors the $c$ axis and competes with
that of the rare earth. For instance in TbCo$_5$, a spin reorientation from
the basal plane to the $c$ axis above 440 K is observed \cite{Lemaire66}. In
the solid solutions R(Ni$_{1\mbox{-}x}$Co$_x$)$_5$, which also crystallize
in the CaCu$_5$ structure, the Co magnetic moment for high $x$ values is
comparable to that in the RCo$_5$ compounds. The Curie temperatures then
progressively decrease when substituting Co by non magnetic Ni ions.

The XMCD measurements at the Co K edge have been performed in LaCo$_5$,
TbCo$_5$, Dy(Ni$_{0.2}$Co$_{0.8}$)$_5$ and TbCo$_2$. Their crystallographic
structures and magnetic properties are reported in table \ref{tab1}. Table 
\ref{tab2} gives the equivalent positions with their coordinates and the
point symmetry group for the CaCu$_5$-type hexagonal structure. All these
compounds were prepared at the Laboratoire de Magn\'etisme Louis-N\'eel, by
RF-melting in a cold crucible from the stoichiometric proportions of pure
elements. In order to minimize oxidation, the melting was performed under
neutral Argon atmosphere. For transmission mode XMCD experiments, the
samples were then crushed into powder and layered onto a thin Kapton foil. The
powders were previously analyzed by X-ray diffraction using a Debye-Scherrer
method. The Co sample consists in a thin metallic foil.

\subsection{XMCD measurements}

The XMCD measurements were carried out at LURE, on the energy dispersive
beam-line D11 \cite{baudeletetal}. The polychromator consists of a curved
Si(111) crystal which focuses the X-ray beam in the middle of the
electromagnet poles. Higher harmonics are rejected by a SiO$_2$ plane mirror
located downstream of the photodiode array detector. Right circularly
polarized X-ray beam is selected by 1 mm-wide slits positioned 3 mrad below
the orbit plane. The circular polarization rate of the incoming radiation is
estimated at $Pc=0.65$. The XMCD spectra are recorded in transmission mode
by selecting one polarization and reversing the magnetic field, applied
along the photon propagation direction. The applied magnetic field intensity
is about 0.4 T. The samples are oriented so that the magnetic field is
applied perpendicular to the plane of the sample. For low temperature
measurements (down to 10 K), the sample can be mounted in a helium
cryogenerator inserted between the electromagnet poles. The spectra have
been measured at room temperature in Co metal, LaCo$_5$ and TbCo$_5$ and at
10 K in TbCo$_2$ and Dy(Ni$_{0.2}$Co$_{0.8}$)$_5$.

\begin{figure}[tbp]
\epsfxsize=16cm
\centerline{\epsfbox{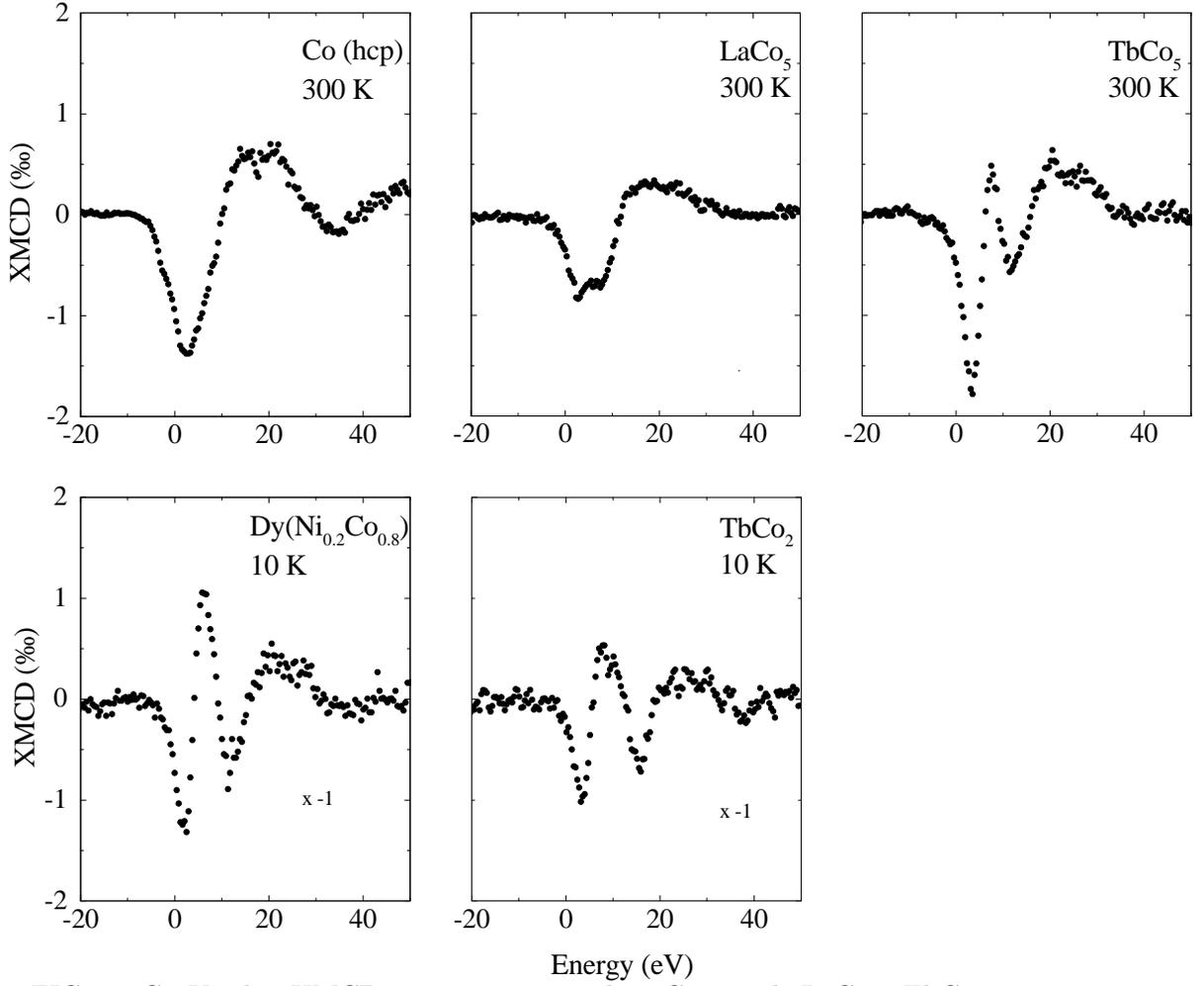}}
\caption{Co K edge XMCD spectra measured in Co metal, LaCo$_5$, TbCo$_5$ at
room temperature and in Dy(Ni$_{0.2}$Co$_{0.8}$)$_5$ and TbCo$_2$ at 10 K.
At room temperature, the magnetization of TbCo$_5$ is dominated by the Co
sublattice like in Co metal and LaCo$_5$. On the contrary at 10 K in
Dy(Ni$_{0.2}$Co$_{0.8}$)$_5$, the magnetization is dominated by the rare earth
sublattice. In TbCo$_2$ the magnetization is dominated by the rare earth
sublattice in the whole ordered range. The spectra of the last two compounds
have been multiplied by -1 to keep the same sign convention than in Co metal.
}
\label{cokexp}
\end{figure}

The XMCD spectra at the Co K edge are presented on figure \ref{cokexp} for
Co metal, LaCo$_5$ and TbCo$_5$ along with other R-Co compounds. The origin
of the energy scale has been chosen at the inflexion point of the absorption
edge. All XMCD spectra have been normalized to the edge jump of the
absorption edge. In TbCo$_2$ and in Dy(Ni$_{0.2}$Co$_{0.8}$)$_5$ at 10 K
($<$T$_{comp}$), the magnetization of the Co sublattice is smaller than the
magnetization of the rare earth sublattice and antiferromagnetically coupled
with it, giving rise to a reverse XMCD signal with respect to Co metal or
LaCo$_5$ and TbCo$_5$ compounds. In order to keep the same sign convention
than in Co metal the XMCD spectra of TbCo$_2$ and
Dy(Ni$_{0.2}$Co$_{0.8}$)$_5$ were multiplied by -1 as indicated on
figure \ref{cokexp}.

The dichroic signal at the Co K edge in Co metal exhibits a one-peak
structure centered about 5 eV above the absorption edge. The full width at
half maximum is about 5 eV. The XMCD spectrum of LaCo$_5$, where the rare
earth is not magnetic, is close to that of Co metal. It presents a negative
one-peak structure at about 5 eV in addition to a weak positive
contribution in the middle of the broad negative structure. On the other
hand, the XMCD spectra obtained from magnetic R-based compounds strongly
contrast with that of Co metal and LaCo$_5$. The dichroic signal consists of
a three-peaks structure, two negative and one positive in the middle of the
negative ones. The central peak exhibits now a huge amplitude, comparable to
that of the negative dips. When switching from an hexagonal-based compound
(TbCo$_5$) to its cubic-based counterpart (TbCo$_2$), sizeable changes in
the XMCD signal are also observed. The central positive feature in TbCo$_2$
seems of larger amplitude and wider than in TbCo$_5$. At higher energy,
all XMCD Co K edge spectra exhibit a positive bump located about 20 eV
above the absorption edge, which may be ascribed to magnetic EXAFS.

The behavior of the XMCD signal at the Co K edge in R-Co compounds, with
respect to Co metal, clearly shows the R influence on the dichroism. In
order to obtain a deeper insight into such a behavior, we performed
multiple-scattering calculations at the Co K edge in Co metal, LaCo$_5$ and
TbCo$_5$.

\section{multiple-scattering calculations at the C\lowercase{o} K edge}

\subsection{ multiple-scattering approach}

The Co K edge XMCD spectra were calculated within the multiple-scattering
framework including the spin-orbit coupling. The cross section is calculated
from the Dirac-Green function according to Eq.\ (3) in Ref.\
\cite{Brouder96}. A fully relativistic core-state wave function has been
used. The determination of the cross section requires the calculation of
the atomic potential which has been carried out within the
self-consistent LMTO approach (see section \ref{sublmto}). The influence of
the relativistic effects, that is the spin-orbit interaction, is obtained by
the series expansion of the Dirac-Green function. To zero-th order in $1/c$
we find the absorption cross section $\sigma_0$ in the absence of
relativistic effects. The expansion carried out up to second order in
$1/c$ provides 5 terms. Only the fourth term of this expansion contributes to
XMCD. This term will be denoted as $\sigma_1$ according to the notation
in Ref.\ \cite{Brouder96}.

In order to take into account the finite core-hole lifetime $\hbar/\Gamma$,
the spectra must be convoluted by a Lorentzian with a half width at half
maximum (HWHM) $\Gamma$. Within the Green function formalism, this
convolution is achieved in calculating the Green function for complex
energies $E+i\Gamma$. We have to consider also the fact that only the
transitions into the empty states, located above the Fermi energy, are
allowed. Thus, the convolution by the $\Gamma$-width Lorentzian must be
carried out from E$_F$ instead of $-\infty$. We will see later on that the
right choice of the Fermi energy is of crucial importance when comparing the
calculated spectra to the experimental ones.

The total absorption cross-section can be written as the sum of two
contributions according to Eq.\ (21) in Ref.\ \cite{Brouder96}.

\begin{equation}
\sigma_0=\sum_s\mbox{Im}\left[\sigma_{0a}^{s}(E+i\Gamma)+
\sigma_{0n}^{s}(E+i\Gamma)\right]
\end{equation}

where the sum is carried out over the two spin states {\it s}. $\sigma_{0a}$
is defined as the {\it atomic} contribution to X-ray absorption and
$\sigma_{0n}$ stands for the influence of the {\it neighbors} surrounding the
absorption site. The atomic contribution $\sigma_{0a}$ is the absorption due
to an isolated atom, and does not exhibit oscillations. The dichroic cross
section defined as $\sigma_{XMCD}=\sigma_1(\epsilon^-)-\sigma_1(\epsilon^+)$
is shown to be equal to the sum of three terms :

\begin{equation}
\sigma_{XMCD}=\mbox{Im}\left[\sum_s(-1)^{(s-1/2)}
\left[\sigma_{1a}^s(E+i\Gamma)+
\sigma_{1l}^s(E+i\Gamma)+\sigma_{1n}^s(E+i\Gamma)
\right]\right]
\end{equation}

where $\sigma_{1a}$ describes the purely {\it atomic} contribution to XMCD
(related to the Fano effect), $\sigma_{1l}$ provides the {\it local}
contribution due to the spin polarization of the $p$-states on the absorbing
site and $\sigma_{1n}$ is the contribution due to the scattering of the
photoelectron by the spin-orbit potential of the {\it neighbors} and the
absorber itself \cite{Note98}. The $p$, $d$, $f$,... orbitals of each atom
of the cluster give a specific contribution to $\sigma_{1n}$. The
expansion of $\sigma_{1n}$ into orbitals and sites is useful to
determine the physical origin of some peaks of the XMCD spectra.

\subsection{calculations of the crystal potential}

\label{sublmto} The calculations were carried out at IDRIS (Orsay). The
crystal potentials were calculated within the local density approximation
using a LMTO basis set. The electronic structure is selfconsistent and the
total energy is converged beyond 0.1 mRy. The spin-orbit coupling is not
included in the calculations. For the Brillouin zone integration of the
density of states  we used the tetrahedron method with about 300 ${\bf k}$-points
in the irreducible part of the Brillouin zone (IBZ) \cite{jepsen72}.

\begin{figure}[tbp]
\epsfxsize=16cm
\centerline{\epsfbox{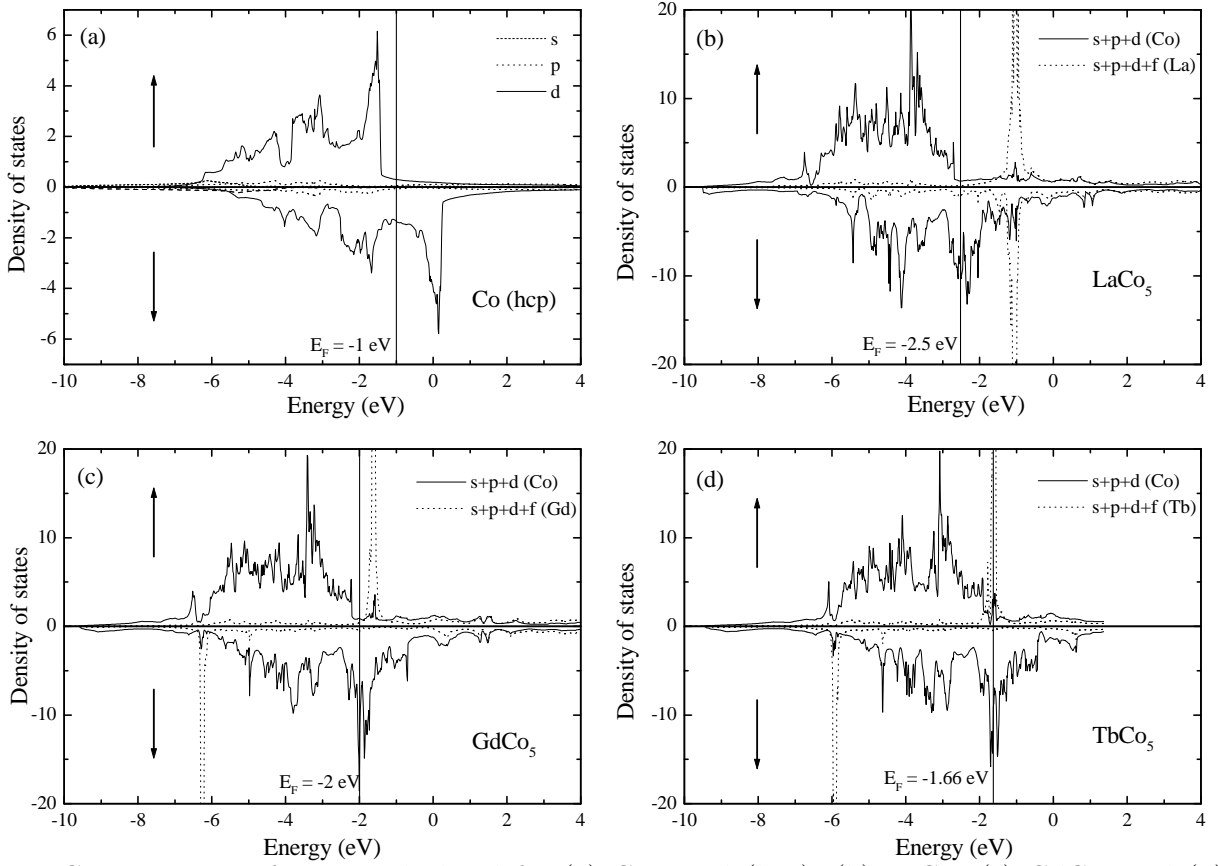}}
\caption{Density of states calculated for (a) Co metal (hcp), (b) LaCo$_5$
(c) GdCo$_5$ and (d) TbCo$_5$. For the RCo$_5$ compounds, the density of
states is drawn so that the Co majority band has the same direction as
the majority band of the Co metal. According to this convention the 4f
occupied states are spin $\downarrow$ polarized.}
\label{dos}
\end{figure}

The calculated Co density of states (DOS) is presented on figure \ref{dos}a.
The result is consistent with previous calculations performed by Jarlborg 
{\it et al.\/} in Co-hcp using the same LMTO formalism \cite{Jarlborg84}.
The similarities between the spin $\uparrow$ and the spin $\downarrow$ DOS
show the rigid band like behavior of Co metal. The calculated Co spin moment
1.57 $\mu_B/$Co agrees well with the experimental value of about 1.6
$\mu_B$. The selfconsistent Fermi level is located at -1 eV.

The calculations on RCo$_5$ class compounds were first performed on
GdCo$_5$. While not measured in the present study, this compound serves as a
reference for our LMTO calculations of the RCo$_5$. The muffin-tin radii R$_{MT}$
of both Gd and Co were derived from the values given by Yamaguchi and
Asano \cite{Yamaguchi94} by expanding the radii so that the total volume
occupied by the muffin-tin spheres in the unit cell is equal to the volume
of the unit cell itself, while keeping the ratio R$_{MT}$(Gd)/R$_{MT}$(Co)
constant. Under this condition, the muffin-tin spheres on two neighboring
sites overlap each other. The R muffin-tin radius in the other RCo$_5$
compounds has been deduced from that of Gd in GdCo$_5$ by scaling the rare
earth radius to the volume of the unit cell, keeping the Co muffin-tin
radius constant throughout the RCo$_5$ series. The muffin-tin radii used for
the calculations are summarized in table \ref{tab3}.

Our calculated GdCo$_{5}$ DOS, presented in figure \ref{dos}c, is comparable
to that of Keller {\it et al.\/} \cite{Keller85}, although the energy
resolution of the latter is not as accurate as in the present study. The
4$f$ spin moment is coupled antiferromagnetically to the mainly
$d$-character moment of the Co conduction band, as expected for rare
earth-transition metal alloys. The Fermi energy is located at E$_{F}$=-2
eV. The calculated spin moments on Co sites, of 1.28 $\mu_B/$Co and 1.37
$\mu_B/$Co respectively on 2c and 3g sites, are slightly smaller than
those previously reported by Yamaguchi and Asano \cite{Yamaguchi94} but
agree with a larger moment on the 3g site. The total spin moment calculated
on the Gd site is 7.06$\mu_B/$Gd, this value is also smaller than the
values reported in Ref. \cite{Yamaguchi94}. The Gd 4$f$ spin moment
deduced from our calculations is 6.69 $\mu_B/$Gd. The reduction of the
4$f$ spin moment compared to the 7 $\mu_{B}$ expected for the free ion
comes from the $f$(Gd)-$d$(Co) hybridization in the band structure calculation.

Figures \ref{dos}b and \ref{dos}d show the calculated DOS of LaCo$_5$ and
TbCo$_5$ respectively. For LaCo$_5$ the Fermi energy is located at
-2.5 eV. The calculated Co spin moments are 1.33 $\mu_B/$Co 
and 1.50 $\mu_B/$Co on the 2c and 3g sites respectively. A weak moment
of 0.4 $\mu_B/$La is also calculated on the La site, which is mainly due
to the 5$d$ states. For TbCo$_5$ the Fermi energy is located at -1.66 eV
and the calculated Co moments are 1.37 $\mu_B/$Co and 1.41 $\mu_B/$Co on
2c and 3g atomic sites, respectively. The values of the Co moments calculated
in LaCo$_5$ and TbCo$_5$ are very consistent with those obtained in
GdCo$_5$. The spin moment calculated at the Tb sites is 5.88 $\mu_B/$Tb.
Compared with the moment calculated on the Gd site in GdCo$_5$, this
value appears quite coherent. Here also the $f$(Tb)-$d$(Co) hybridization
leads to a reduction of the 4$f$ spin moment which reaches 5.63 $\mu_B/$Tb.

\subsection{results of multiple-scattering}

The XMCD calculations have been carried out with the numerical program
described in Ref.\ \cite{Brouder96}. The converged potential included in the
multiple-scattering calculations are those calculated by LMTO. The XMCD
program uses touching muffin-tin spheres without overlap. The absence of
overlap is required for the multiple-scattering calculations to be
convergent. Because of the difference in the muffin-tin radii between LMTO
and multiple-scattering potentials, there is an  interstitial
volume between the spheres in the multiple-scattering calculations. This
volume is represented by a muffin-tin potential V$_I$ which is calculated
as the average potential, in the interstitial region, due to all the
charges in the cluster (including the interstitial region). This
interstitial potential V$_I$ is the energy origin of the
multiple-scattering calculations, and has no reason to be the same as
the energy origin of the LMTO calculations. This ambiguity can be
avoided by using full potential codes that do not make the muffin-tin
approximation. However, no such codes are available at the moment for XMCD.

The muffin-tin radii used in the multiple-scattering calculations are
presented on table \ref{tab3}. For LaCo$_5$ and TbCo$_5$, they have been
deduced using the procedure proposed by Yamaguchi and Asano for GdCo$_5$ 
\cite{Yamaguchi94}. Note that the R$_{MT}$(R)/R$_{MT}$(Co) ratio has the
same value than in LMTO calculations for TbCo$_5$ while it is slightly
smaller in LaCo$_5$.

\subsubsection{Co metal}

\begin{figure}[tbp]
\epsfxsize=12cm
\centerline{\epsfbox{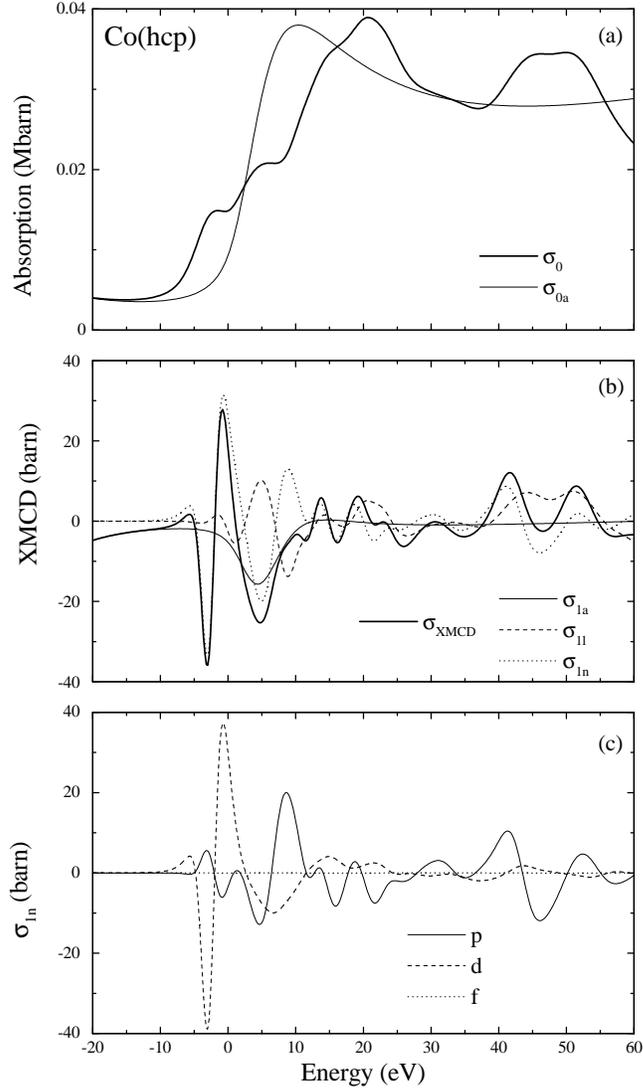}}
\caption{Absorption and dichroic spectra at the Co K edge in Co metal
calculated for a cluster of 105 atoms without introducing the Fermi
level. (a) $\sigma_{0}$ represents the total absorption cross section
and $\sigma_{0a}$ the atomic contribution of the isolated atom. (b)
$\sigma_{XMCD}$ represents the total dichroic spectrum. $\sigma_{1a}$
describes the purely atomic contribution, $\sigma_{1l}$ the contribution
due to the spin polarization of the $p$-states on the absorbing
site and $\sigma_{1n}$ the contribution arising from the scattering of
the photoelectron by the spin-orbit potential of the neighbors. (c) Expansion
of the $\sigma_{1n}$ cross section into the $p$, $d$ and $f$ orbitals of
the first shells of the cluster.}
\label{cocalc}
\end{figure}

The calculations for hcp Co metal have been performed using
a cluster of 105 atoms. The absorption, $\sigma_{0}$, and dichroic,
$\sigma_{XMCD}$, spectra calculated first without introducing the Fermi
level are presented in figures \ref{cocalc}a and \ref{cocalc}b. Figure
\ref{cocalc}b also presents the contribution of $\sigma_{1a}$,
$\sigma_{1l}$ and $\sigma_{1n}$ to the dichroic cross section. The
XMCD spectrum consists in a three-peaks structure near the absorption edge,
followed by weak EXAFS oscillations appearing about 10 eV above the edge.
The first two peaks in the XMCD spectrum come mainly from the
$\sigma_{1n}$ contribution. As aforesaid, this contribution comes from
the scattering due to the spin-orbit potential on the neighbors and the absorbing
site. Quite similar results have been previously reported at the Fe K
edge \cite{Brouder96}. The third peak at 5 eV results from the sum of
the three contributions, $\sigma_{1a}$, $\sigma_{1l}$ and $\sigma_{1n}$
which have comparable amplitudes. As we shall see later in LaCo$_5$ and
TbCo$_5$, in this energy region $\sigma_{1a}$ and $\sigma_{1n}$ are
always in phase opposition with $\sigma_{1l}$. $\sigma_{1l}$ originates form
the polarization of the $p$-states on the absorbing site. The expansion of
the $\sigma_{1n}$ cross section into the $\ell$ components and sites of
the whole cluster is presented on figure \ref{cocalc}c. The scattering due
to the $d$ spin-orbit potential mainly accounts for the first two peaks
in the near-edge region of the XMCD signal. The amplitude of the
oscillations due to the scattering by the spin-orbit potential on the
$p$ orbitals starts to significantly increase above the edge and a large
peak is calculated around 8 eV. The negative peak around 5 eV in the
$\sigma_{1n}$ cross section results from the well balanced contributions
of the $p$ and $d$ orbitals.

\begin{figure}[tbp]
\epsfxsize=12cm
\centerline{\epsfbox{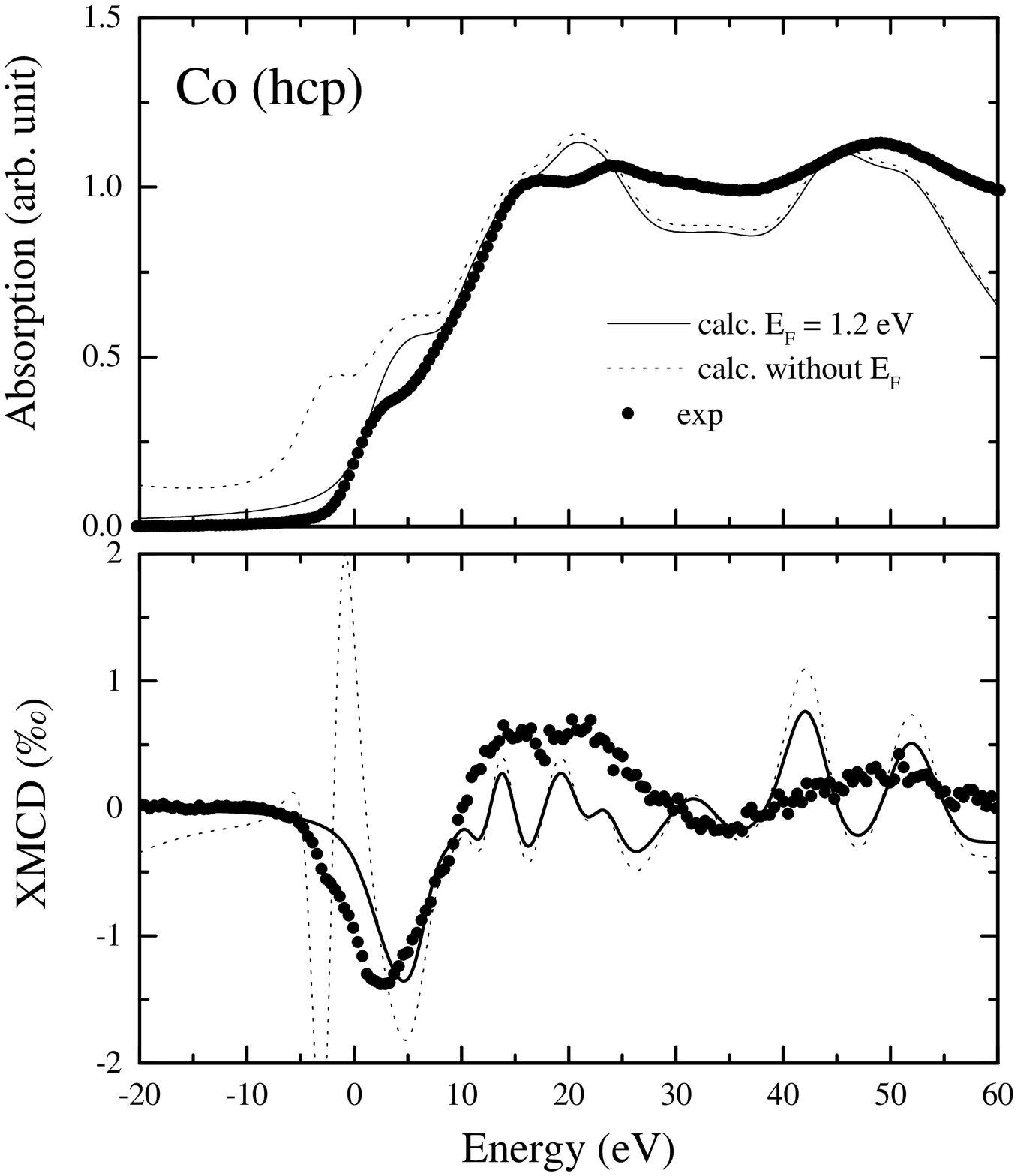}}
\caption{Absorption and XMCD experimental (solid circle) and calculated (line)
spectra at the Co K edge in Co metal. The calculated spectra are normalized to
the edge jump at the absorption edge. The solid lines represent the spectra
calculated with E$_F$=1.2 eV and the dotted lines the spectra calculated
without the Fermi energy.}
\label{cocalcexp}
\end{figure}

In order to compare the calculated spectra to the experimental ones, the
Fermi energy has to be included. In figure \ref{cocalcexp} the absorption
and dichroic experimental spectra are presented along with the calculated
ones with and without introducing the Fermi energy. Absorption and
 XMCD spectra are all normalized to the edge jump at the absorption
edge so that the absorption edge jump is 1. The Fermi energy was
chosen at 1.2 eV by fitting both the absorption and the XMCD spectra to
the experimental ones in the near-edge region. The Fermi energy of the
multiple-scattering calculations is not the same as that of the LMTO
calculations because of the difference in the energy origin between the
two approaches. The effect of the Fermi energy clearly shows up, especially
for the XMCD spectrum : the first two peaks have vanished and only
subsists the negative peak around 5 eV. The XMCD spectrum is
broadened around E$_F$ as expected from the truncated Lorentzian
convolution. It may be noticed that in Ref.\ \cite{Brouder96}, the XMCD
spectrum calculated at the K edge in Fe metal with E$_F$=0 eV keeps a
positive peak at low energy, mainly coming from the $\sigma_{1n}$
contribution, which fits well the derivative-like behavior of
the measured XMCD spectrum. The higher Fermi level found in Co metal
leads to the collapse of this second peaks in the near-edge region of
the Co K edge XMCD spectrum. It is worth noting that a higher Fermi
level in Co metal is coherent with the higher occupancy of the Co $d$ band
($3d^7$) compared to Fe ($3d^6$).

\begin{figure}[tbp]
\epsfxsize=12cm
\centerline{\epsfbox{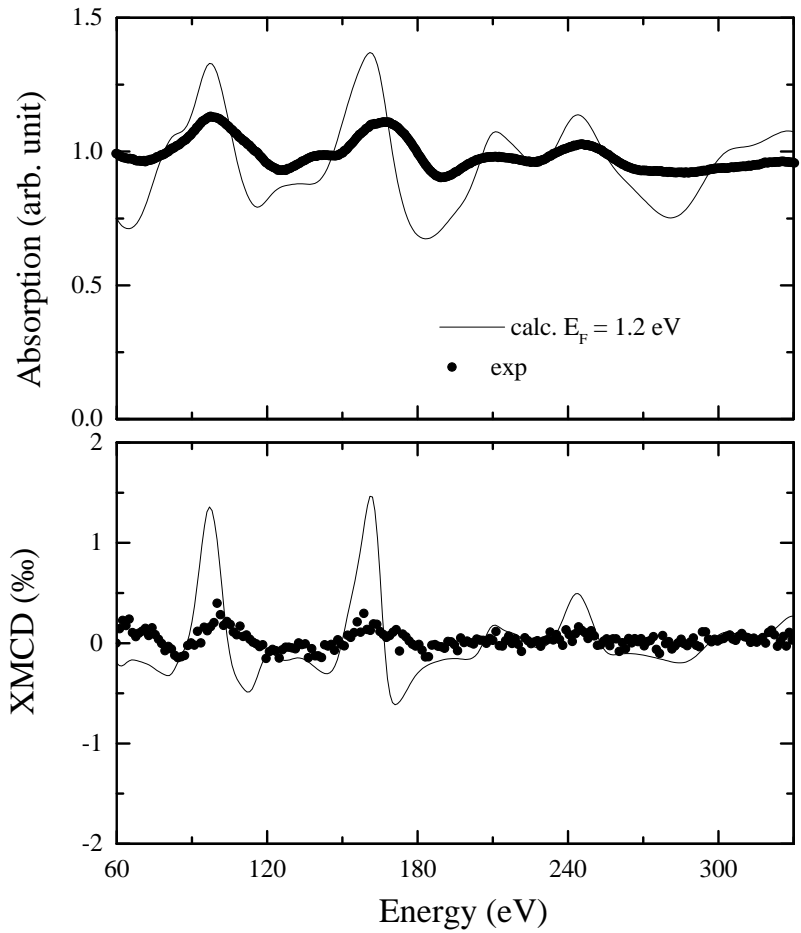}}
\caption{High energy part of the normalized absorption and XMCD spectra at
the Co K edge in Co metal. The solid circles represent the experimental
spectra and the solid lines the spectra calculated with E$_F$=1.2 eV.}
\label{coexafs}
\end{figure}

In the near edge region the calculation reproduces rather well the
structure of the absorption and XMCD experimental spectra. The
calculated spectra present a slight shift to the high energies. This
shift is very likely due to the difficulty to properly fit in the energy
between calculated and experimental spectra. For the XMCD spectrum the
calculation yields the expected one-peak feature with however a narrower width
than the experimental one. In the region between 10 eV and 30 eV the
calculated oscillations are not resolved in the experimental spectrum,
where only a large bump is observed. The extended energy spectra are
presented on figure \ref{coexafs}. The slight shift in energy of the
EXAFS oscillations is again noticed between experiment and calculation in
both the absorption and the XMCD spectra. Though the EXAFS structures
are strongly damped in the measured spectra, the agreement between
experiment and calculation remains rather satisfactory. The agreement in
the intensity of the calculated and experimental peak around 5 eV is
very likely fortuitous. Indeed in the calculation the degree of circular
polarization is set equal to 1, while experimentally the polarization
rate of the incoming beam, estimated at 0.65, reduces the signal
intensity. This is confirmed in the energy range 10-30 eV where the
intensity of the calculated oscillations is to small to account for the
bump observed experimentally. Consequently it turns out that in the high
energy region the calculated amplitude of the oscillations is always
overestimated, in the dichroic spectrum but also in the absorption one.
These effects might be related to a current limitation of the XMCD program
which does not include the inelastic scattering processes undergone by
the photoelectron in the metal.

\subsubsection{RCo$_5$}

The multiple-scattering calculations in the LaCo$_5$ and TbCo$_5$ compounds
have been carried out on a 117 atoms cluster for each Co atomic site. The
total spectrum results from the weighted sum of the spectrum on each site with
respect to its multiplicity in the unit cell. Calculations have been performed
only in the near edge energy domain since the extended structures are poorly
resolved in the absorption and dichroic experimental spectra. 

\begin{figure}[tbp]
\epsfxsize=12cm
\centerline{\epsfbox{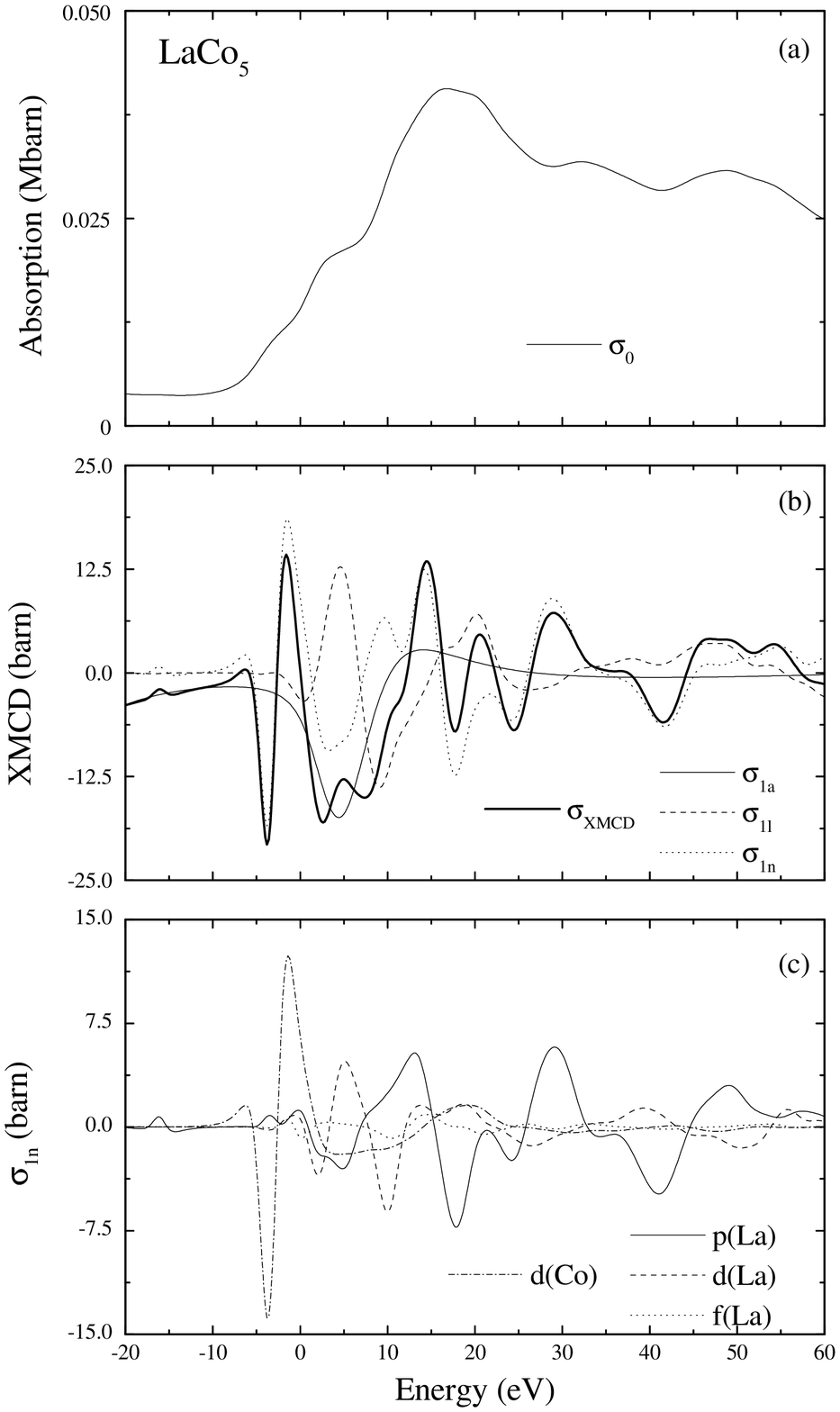}}
\caption{Absorption and dichroic spectra at the Co K edge in LaCo$_5$
calculated without introducing the Fermi level. The multiple-scattering
calculations have been carried out on a 117 atoms cluster for each Co atomic
site. The total spectrum results from the weighted sum of the spectrum on
each site with respect to its multiplicity in the unit cell. (a) $\sigma_{0}$
represents the total absorption cross section. (b) $\sigma_{XMCD}$ represents
the total dichroic spectrum. $\sigma_{1a}$ describes the purely atomic
contribution to the dichroic spectrum, $\sigma_{1l}$ the
contribution due to the spin polarization of the $p$-states on the absorbing
site and $\sigma_{1n}$ the contribution of the spin-orbit potential of the
neighbors. (c) Expansion of $\sigma_{1n}$ cross section into the $p$, $d$ and
$f$ orbitals of the La first neighbors and into the $d$ orbitals of the
Co first neighbors.}
\label{laco5calc}
\end{figure}

\begin{figure}[tbp]
\epsfxsize=12cm
\centerline{\epsfbox{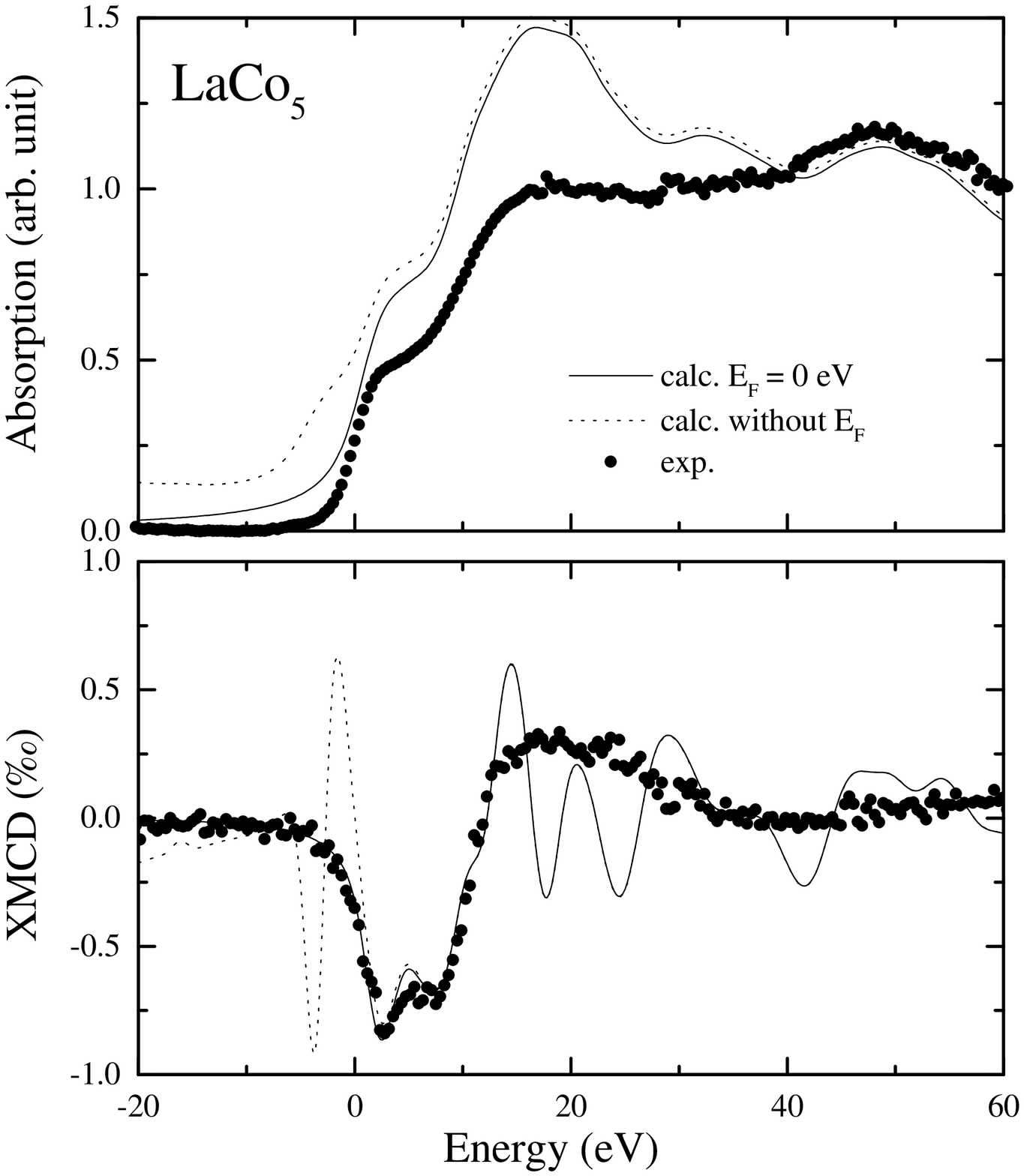}}
\caption{Absorption and dichroic experimental (solid circle) and
calculated (line) spectra at the Co K edge in LaCo$_5$. Calculated spectra
are normalized to the edge jump at the absorption edge. The solid lines
represent the spectra calculated with E$_F$=0 eV and the dotted lines
the spectra calculated without the Fermi energy.}
\label{laco5calcexp}
\end{figure}

The absorption and dichroic spectra calculated without the Fermi level for
LaCo$_5$ are presented on figures \ref{laco5calc}a and \ref{laco5calc}b. The
different components, $\sigma_{1a}$, $\sigma_{1l}$ and $\sigma_{1n}$, to the
XMCD cross section are also shown. In the energy region -10 to 0 eV, the
$\sigma_{1n}$ contribution is responsible for the first two peaks as
already observed in Co metal. Around the third negative
peak at 5 eV, the $\sigma_{1a}$ and $\sigma_{1l}$ contributions are
comparable to those calculated in Co metal whereas $\sigma_{1n}$ is reduced
by more than a factor 2. The reduction of $\sigma_{1n}$ and the phase
opposition between $\sigma_{1a}$, $\sigma_{1n}$ and
$\sigma_{1l}$ give rise to the positive cusp located
in the middle of the negative peak, characteristic of the LaCo$_5$ XMCD
spectrum (cf. fig.\ \ref{laco5calcexp}). Figure \ref{laco5calc}c
presents the contribution to the $\sigma_{1n}$ cross section of the $p$,
$d$ and $f$ shells of the La first neighbors and of the $d$ shell of the
Co first neighbors. In the near-edge region and just above the edge,
$\sigma_{1n}$ is dominated by the scattering processes due to the $d$
contribution of Co and La. The peculiar decrease of the total
$\sigma_{1n}$ cross section around 5 eV comes from the existence
of a positive peak in the La $d$ orbital component, while the Co $d$ and La
$p$ components give rise to negative structures. At higher energies
features on $\sigma_{1n}$ come essentially from the La $p$
shell contribution.

The normalized absorption and XMCD cross sections calculated with a
Fermi energy of 0 eV are compared on figure \ref{laco5calcexp} with the
experimental spectra. The calculated dichroic spectrum was divided by an
arbitrary factor in order to fit the experimental one. Besides the reduction
of the signal due to the circular polarization rate, this correction
accounts for the fact that the magnetic saturation of the powdered sample
of LaCo$_5$ is not reached under an applied field of 0.4 T. The agreement
for the absorption spectrum is far from being excellent but the dichroic
signal is well reproduced, especially the structure at 5 eV. It can be
noticed that the total width of the spectrum agrees well with the observed
one. Above 10 eV discrepancies between the calculations and the measured
spectrum are observed, as for Co metal the experimental signal presents a
large bump instead of well resolved structures.\newline

\begin{figure}[tbp]
\epsfxsize=12cm
\centerline{\epsfbox{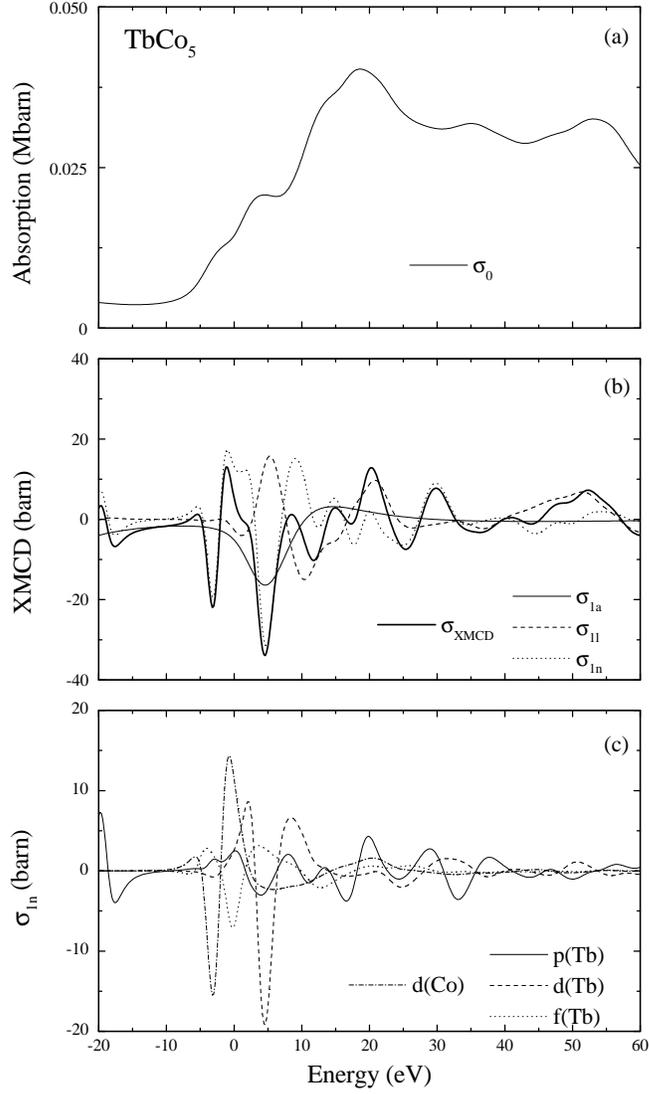}}
\caption{Absorption and dichroic spectra at the Co K edge in TbCo$_5$
calculated without introducing the Fermi level. Multiple-scattering
calculations have been performed on a 117 atoms cluster for each Co atomic
site. The total spectrum is the weighted sum of the spectrum on each site
with respect to its multiplicity in the unit cell. (a) $\sigma_{0}$
represents the total absorption cross section. (b) $\sigma_{XMCD}$ is the
total dichroic spectrum. $\sigma_{1a}$ describes the purely atomic
contribution, $\sigma_{1l}$ the contribution due to the spin polarization
of the $p$-states on the absorbing site and $\sigma_{1n}$ the contribution due
to the scattering of the photoelectron by the spin-orbit potential of
the neighbors. (c) Expansion of $\sigma_{1n}$ into the $p$, $d$ and $f$
orbitals of the Tb first neighbors and $d$ orbitals of the Co first neighbors.}
\label{tbco5calc}
\end{figure}

\begin{figure}[tbp]
\epsfxsize=12cm
\centerline{\epsfbox{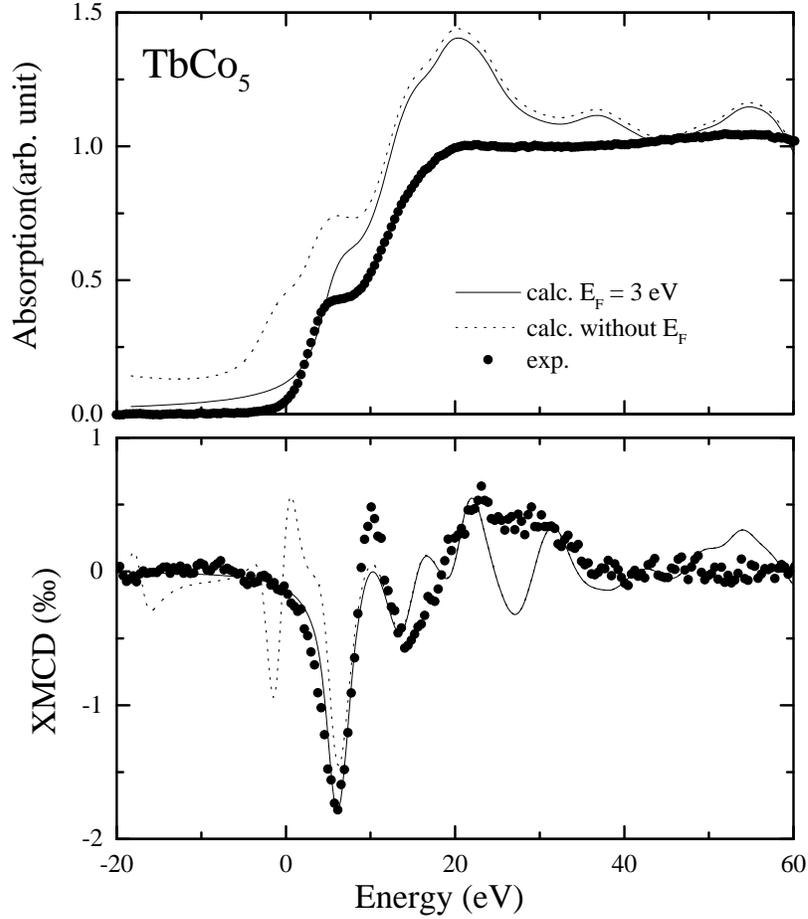}}
\caption{Absorption and dichroic 
spectra at the Co K edge in TbCo$_5$. Solid circles represent the experimental
data. Solid lines represent the spectra calculated with E$_F$=3 eV and
dotted lines the spectra calculated without the Fermi energy. The
calculated spectra are normalized to the edge jump at the absorption edge.}
\label{tbco5calcexp}
\end{figure}

The absorption and XMCD calculated spectra without the Fermi level for TbCo$_5$
are illustrated on figure \ref{tbco5calc}a and \ref{tbco5calc}b. Figure
\ref{tbco5calc}b also shows the contributions of $\sigma_{1a}$,
$\sigma_{1l}$ and $\sigma_{1n}$ to the total XMCD cross section.
The structures of the XMCD signal in the low energy region arise, as in
LaCo$_5$, from $\sigma_{1n}$. The first negative peak has a structure and
an amplitude very similar to those obtained in LaCo$_5$ or Co metal. The
second positive peak presents a double structure which results from the mixing
of the Co $d$ and Tb $d$ shell positive contributions and the negative Tb
$f$ shell contribution as shown in figure \ref{tbco5calc}c. In figure
\ref{tbco5calc}c are represented only the contributions to the
$\sigma_{1n}$ cross section from the $p$, $d$ and $f$ shells of the Tb
first neighbors and the $d$ shell of the Co first neighbors. The addition
of the contributions, of smaller intensity, coming from the other atoms
of the cluster does not change quantitatively the structure of the
spectra. It may be noted that the Tb $f$ shell contribution is not
negligible in the near edge region. Compared to the LaCo$_5$ XMCD
spectra, the third negative peak around 5 eV comes almost exclusively
from $\sigma_{1n}$. In this energy region $\sigma_{1a}$ and $\sigma_{1l}$
having roughly the same amplitude and opposite sign cancel each other.
This negative structure can be mainly assigned to the Tb $d$ shell
contribution. This contribution along with the Tb $p$ and $f$
contributions, of lower intensity, give rise also to the
positive bump in $\sigma_{1n}$ which immediately follows the negative
peak. When the Fermi level is included in the calculation, the
structures near the absorption edge are smeared out. The calculated spectra
which best fit with the experimental ones are obtained with E$_F$=3 eV.
Experimental and calculated spectra are presented in figure
\ref{tbco5calcexp}. The calculated absorption spectrum presents structures
which are not resolved in the experimental one. The calculated dichroic
spectrum was scaled to the experimental one. As for LaCo$_5$, this
correction accounts for the non-saturation of the magnetization. The two
peaks at 6 and 10 eV in the experimental XMCD spectrum are quite well
reproduced by the calculations, in particular their widths. At
higher energies, we find incoherences again between the calculation and
the measured spectrum.

\section{Concluding Remarks}

\begin{figure}[tbp]
\epsfxsize=16cm
\centerline{\epsfbox{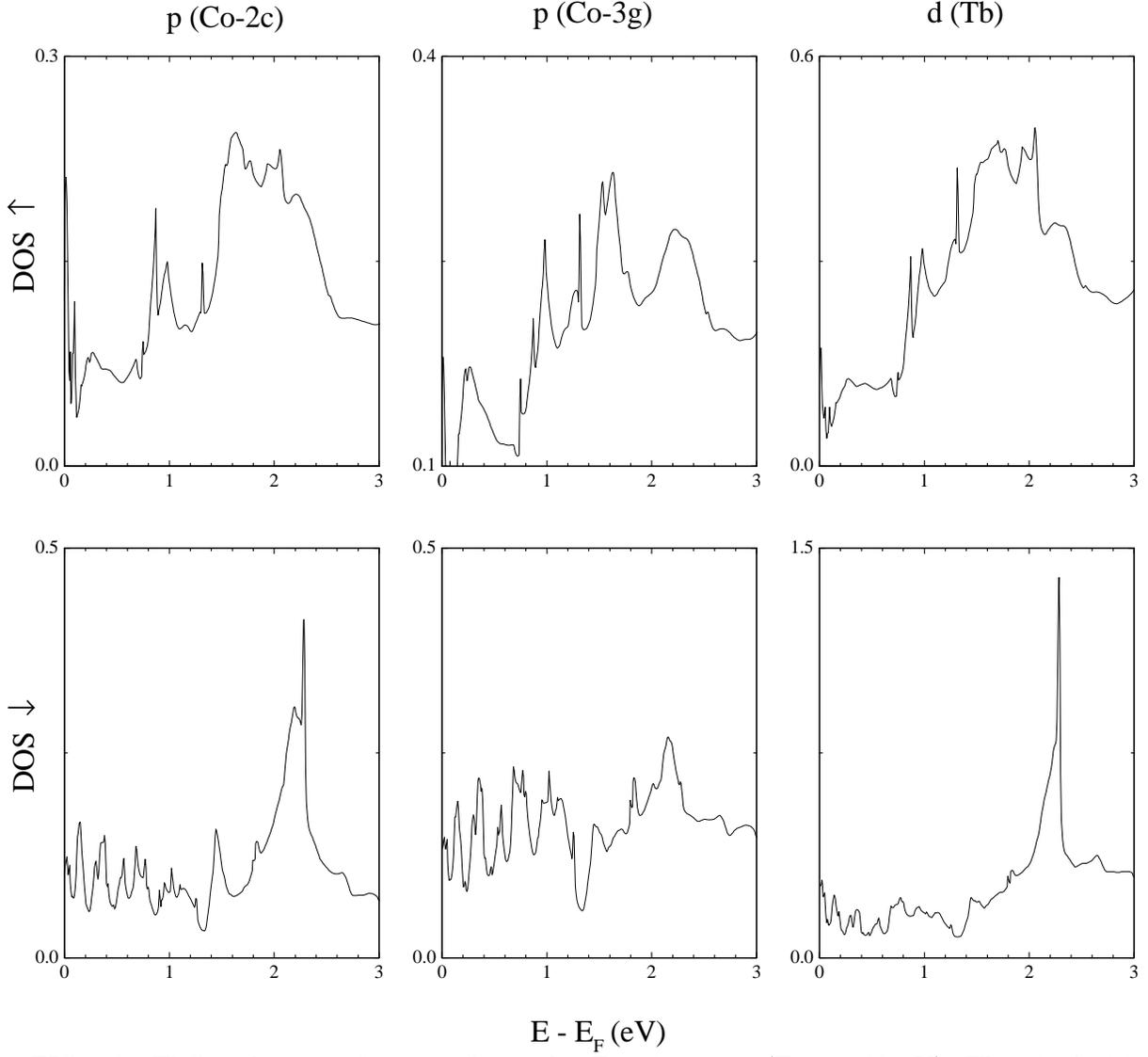}}
\caption{TbCo$_5$ density of states above the Fermi energy (E$_F$=-1.66 eV).
The similarities between the $d$(Tb) and the $p$(Co) densities of states
emphasize the strong influence of the rare earth on the polarization of Co
in the RCo$_5$ compounds. The closer resemblance with the $p$(Co-2c) is due to
the shorter Tb-Co(2c) distance.}
\label{tbco5doszoom}
\end{figure}

Calculations within the multiple scattering approach of the dichroic signal
at the Co K edge in pure Co metal reproduce with a rather good agreement
the structures of the experimental spectrum. In particular the one-peak
structure near the edge is well accounted for with a Fermi energy of 1.2
eV. The Co K edge spectrum, calculated without the Fermi level presents
strong similarities with the Fe K edge spectrum calculated in the same
conditions. In the present theoretical approach it is possible to
separate the local contributions to the dichroic cross section from
those coming from the surrounding atoms. When these contributions are
examined it is confirmed that the scattering by the spin-orbit potential
of the $d$ shell of the neighboring atoms creates strong structures in
the low energy range of the spectrum. This is in agreement with Igarashi
and Hirai calculations. The local contributions, atomic contribution and
contribution due to the spin polarization of the $p$-states, give structures
at higher energies typically around 5 eV. However as soon as the Fermi level
is introduced in the calculations, low energy structures partially (Fe),
or completely (Co), disappear. Present calculations show that the
negative structure observed in the Co metal XMCD spectrum at the Co K edge
results from a well balanced mixing between local, $\sigma_{1a}$ and
$\sigma_{1l}$, and surrounding, $\sigma_{1n}$, contribution. It may be noted
that, in this energy range, the contributions to $\sigma_{1n}$, arising from
the spin-orbit coupling in the $d$ and $p$ shells have comparable intensities.
The calculated Co K edge XMCD spectra in LaCo$_5$, and TbCo$_5$
reproduce, with an excellent agreement in the near edge region, the structures of
the experimental spectra. As for Co metal the signal structure in LaCo$_5$ is
accounted for by a balanced mixing between local and surrounding
contributions. In this last contribution the rare earth atoms, via the
spin-orbit coupling on the La $d$ and $p$ shells, significantly influence the
structure. In the case of TbCo$_5$ the multiple-scattering calculations
stress the major role of the contributions due to the neighbors, $\sigma_{1n}$
in the structure of the spectrum around 5 eV. More precisely, the expansion
of the $\sigma_{1n}$ cross section into the $\ell$ orbital components allowed us
to evidence that the first negative peak in the XMCD spectrum at the Co K
edge originates almost exclusively from the spin-orbit potential on
the Tb $d$ states. This result points out that the structures of the XMCD
spectrum at the K edge of the transition metal in the R-TM alloys are induced to
a large extent by the rare earth when the rare earth is magnetic.
The XMCD at the Co K edge in the R-Co compounds, which probes the empty
$p$-states on Co sites, thus detects the spin-orbit interaction on the rare
earth $d$-states. This process is very likely mediated through the
hybridization. The similarities observed in the density of states above
the Fermi level, between the Tb($d$) and the Co($p$)-DOS in TbCo$_5$ (cf.
fig.\ \ref{tbco5doszoom}) underline the existence of a Co($p$)-R($d$)
hybridization. The closer resemblance between Co-2c($p$) and Tb($d$) is
related to the shorter distance leading to a larger orbital overlapping
between the Co-2c--Tb compared to Co-3g--Tb.

Effects of the environment can be expected to exist at the L$_{2,3}$ 
edges of the transition metal as well, like in the RE-TM alloys where the
RE(5$d$) and TM(3$d$) band are hybridized. For instance, it is very likely
that magnetic EXAFS structure can reflect these effects and recent
theoretical investigations on the magnetic EXAFS at the L$_{2,3}$ edges
of pure Fe are very promising \cite{Ebert98}. However a  proper
experimental evidence presents great difficulties as the white lines dominate
the spectra at these edges. Moreover, in the soft x-ray energy range, surface
defaults which are far from being negligible in the RE-TM alloys may
disturb the interpretation of the spectra.

Though the XMCD spectra are quite well reproduced in the near region, a
better agreement has yet to be obtained in the EXAFS
region both in the absorption and dichroic spectra. An 
improvement planned for the future is the use of an optical
potential to take into account the exchange and inelastic interactions on
the photoelectron.

{\bf Acknowledgments}  M.A. acknowledges support in part by the U.S.
Department of Energy Basic Energy Sciences, Division of Materials Sciences
and by NSF, grant number DMR-9520319.

\newpage

\begin{table}[tbp]
\begin{tabular}{cccccc}
Compound & Structure & a(\AA) & c(\AA) & T$_C$(K) & T$_{comp}$(K) \\ \hline
LaCo$_5$ & CaCu$_5$ & 5.105 & 3.966 & 840 & - \\ 
TbCo$_5$ & CaCu$_5$ & 4.950 & 3.979 & 980 & 90-110 \\ 
Dy(Ni$_{0.2}$Co$_{0.8}$)$_5$ & CaCu$_5$ & 4.90 & 4.00 & $\approx$1000 & $\approx$200
\\ 
TbCo$_2$ & MgCu$_2$ & 7.206 & - & 237.5 & -
\end{tabular}
\caption{Crystallographic and magnetic data of the studied RCo$_x$ compounds.
}
\label{tab1}
\end{table}

\newpage 
\begin{table}[tbp]
\begin{tabular}{cccc}
atoms & sites & point symmetry & equivalent positions \\ \hline
R & 1a & $\frac{6}{m}mm$ & (0,0,0) \\ 
Co & 2c & $\overline{6}m2$ & (1/3,2/3,0), (2/3,1/3,0) \\ 
Co & 3g & $mmm$ & (1/2,0,1/2), (0,1/2,1/2), (1/2,1/2,1/2)
\end{tabular}
\caption{Equivalent positions, point symmetry groups and coordinates of the
equivalent positions for the CaCu$_5$-type hexagonal structure.}
\label{tab2}
\end{table}

\newpage

\begin{table}[tbp]
\begin{tabular}{ccc}
& R$_{MT}$ (LMTO) (\AA) & R$_{MT}$ (MS) (\AA) \\ \hline
Co metal & 1.38 & 1.25 \\ \hline
GdCo$_5$ &  &  \\ \hline
Gd & 1.86 & - \\ 
Co(2c) and Co(3g) & 1.40 & - \\ \hline
LaCo$_5$ &  &  \\ \hline
La & 1.97 & 1.65 \\ 
Co(2c) and Co(3g) & 1.40 & 1.24 \\ \hline
TbCo$_5$ &  &  \\ \hline
Tb & 1.85 & 1.63 \\ 
Co(2c) and Co(3g) & 1.40 & 1.23
\end{tabular}
\caption{Muffin-tin radii used in the calculations. Column 2 gives the
values of the muffin-tin radii, R$_{MT}$(LMTO), adjusted for the LMTO
calculations with overlapping muffin-tin spheres. Column 3 gives the values
of the muffin-tin radii, R$_{MT}$(MS) introduced for the multiple-scattering
calculations with touching spheres (from Ref.\ \protect\cite{Yamaguchi94}).}
\label{tab3}
\end{table}

\newpage

\end{document}